\documentclass[aps,prl,twocolumn]{revtex4}
\usepackage{amsmath}
\usepackage{amssymb}
\usepackage{graphics}

\newcommand\expect[1]{\left<#1\right>}

\newcommand\Ev{\mathbb{E}}
\newcommand\dg{^\dagger}

\newcommand\Tr{\mathrm{Tr}}

\newcommand\minihalf{{\scriptstyle{\frac{1}{2}}}}

\newcommand\Lc{\mathcal{L}}
\newcommand\Fa{F_\alpha}
\newcommand\Fad{F_\alpha\dg}
\newcommand\Hfr{H_\mathrm{fr}}

\begin{document}
\title{New results unearthed from urtexts of quantum state diffusion}  
\author{Lajos Di\'osi}
\email{diosi.lajos@wigner.mta.hu}
\homepage{www.rmki.kfki.hu/~diosi}
\affiliation{Wigner Research Centre for Physics, 
H-1525 Budapest 114. P.O.Box 49, Hungary}
\date{\today}
\begin{abstract}
Thirty years ago, the present author discussed pure state unraveling
(stochastic quantum trajectories) of Markovian open system dynamics.
The fact that he considered all positive dynamics, not restricted
for the Lindblad-Gorini-Kossakowski-Sudarshan complete-positive subclass,
has remained unnoticed so far. Here we point out the importance of the transition
rate operator and the merit of invariant (representation-independent)
approach, with reference to invariant condition of positive
dynamics and extension of  quantum state diffusion beyond
complete-positive dynamics.  A new result, description of all diffusive unravelings of positive dynamics directly follows. 
\end{abstract}
\maketitle
\emph{Introduction.} 
In 1986 and 1988 the present author published two papers \cite{Dio86,Dio88} on what later became known as stochastic quantum trajectories \cite{QT} and quantum state diffusion \cite{QSD}, now standard methods for open quantum systems \cite{QTQSD}. 
Both papers considered the master equation for the density operator 
$\rho$ of Markovian open quantum systems:
\begin{equation}\label{ME}
\dot\rho=\Lc\rho,
\end{equation}
requiring \emph{positive} dynamics, i.e., that the superoperator $\Lc$ conserve the
positivity of $\rho$. 
Then stochastic Schr\"odinger equations (SSEs) were constructed to generate pure state
solutions (quantum trajectories) $\psi$, constituting so-called stochastic \emph{unraveling}  
the dynamics \eqref{ME}. That means the stochastic average of quantum trajectories must yield the ensemble density operator
\begin{equation}\label{unrav}
\Ev\psi\psi\dg=\rho
\end{equation}
which is the solution of the master equation \eqref{ME}. 

Seminal works of Lindblad \cite{Lin76} and Gorini et al. \cite{Goretal76} 
were mentioned but dynamics was not restricted for complete-positive (CP) ones.
Hence the Lindblad-Gorini-Kossakowski-Sudarshan  structure
\begin{equation}\label{MECP}
\dot\rho=-i[H,\rho]+\Fa\rho\Fad-\minihalf\{\Fad\Fa,\rho\}
\end{equation}
was not assumed. (Here and henceforth Einstein convention of summation for repeated indices is used.)  At all, no particular representation of the master
equation \eqref{ME} was introduced. All results were derived and explained in terms of the superoperator $\Lc$, all results were representation-independent, i.e:
\emph{invariant}. All results were valid for positive, not necessarily CP dynamics \eqref{ME}.

Since typical Markovian open quantum systems satisfy the CP master equation \eqref{MECP},
all standard works on quantum trajectories in general and on quantum state diffusion (QSD) \cite{QT,QSD,QTQSD} in particular
have imposed the  structure \eqref{MECP}. Perhaps the only exception was
Gisin's paper in 1990 \cite{Gis90} which determined all diffusive 
quantum trajectories in for all positive 2D Markovian dynamics, including the non-CP ones.
Jump unraveling of a non-CP master equation in 2D appeared next in \cite{Dio14} only.  
Very recently a detailed work \cite{CaiSmiBas16} has
extended QSD for non-CP master equations.

Below we are going to 
recapitulate the cornerstones of the urtexts \cite{Dio86,Dio88}, to 
emphasize the merit of the invariant approach and to unearth the unnoticed 
results. 

\emph{Positive dynamics.}
Ref. \cite{Dio86} \emph{understands} that the conservation of $\rho$'s positivity
is guaranteed if it holds for any pure initial state differentially in time. 
Hence in infinitesimal time $dt$ any pure state must evolve into a non-negative density matrix:
\begin{equation}\label{posdt}
0\leq\psi\psi\dg+dt\Lc(\psi\psi\dg)\equiv\psi\psi\dg+dt L,
\end{equation}
where $L=\Lc(\psi\psi\dg)$ is a useful shorthand notation.
Central object, related to the superoperator $\Lc$, is the
\emph{transition rate operator}:
\begin{equation}\label{W}
W=L-\{L,\psi\psi\dg\}+\expect{L}\psi\psi\dg,
\end{equation}
with notation $\expect{L}=\psi\dg L\psi$.   
The other central object is
the \emph{frictional Hamiltonian} satisfying
 \begin{equation}\label{Hfr}
-i\Hfr\psi=(L-\expect{L})\psi.
\end{equation}
$\Hfr$ is non-linear and non-Hermitian but norm-conserving Hamiltonian. The term to
ensure normalization coincides with the total transition rate 
$w=\Tr W=-\expect{L}$, as it follows from eq. \eqref{W}. 
By substituting eqs. (\ref{W},\ref{Hfr}) in the r.h.s.
of the master equation \eqref{ME}, 
the time-derivative of an initial pure state density operator $\rho=\psi\psi\dg$
takes the form
\begin{equation}\label{MEHfrW}
\dot\rho=-i\Hfr\rho+i\rho\Hfr\dg+W-w\rho.
\end{equation}
Accordingly, the inequality \eqref{posdt} takes this form:
\begin{equation}\label{posdtHfrW}
0\leq\psi\psi\dg-i\Hfr\psi\psi\dg dt+i\psi\psi\dg\Hfr\dg dt + Wdt-w\psi\psi\dg dt.
\end{equation}
Since $W\psi=\psi\dg W=0$ by construction \eqref{W}, the above inequality
is equivalent with the non-negativity of the transition rate operator:
\begin{equation}\label{Wpos}
W\geq0,
\end{equation}
which is \emph{understood} in \cite{Dio86} --- where non-negativity of $W$ is taken
for granted --- as the necessary and sufficient condition on the superoperator $\Lc$ 
to  conserve the positivity of $\rho$. This theorem was explicitly stated and 
re-derived in \cite{CaiSmiBas16},
starting from the oldest conditions of Kossakowski \cite{Kos72} which \cite{Dio88}
quoted in the following invariant form. The dynamics \eqref{ME} is positive if  
the operator $L=\Lc(\psi\psi\dg)$ satisfies
\begin{equation}\label{MEpos}
\psi\dg L\psi\leq0,~~~\psi_\bot L\psi_\bot\geq0
\end{equation}
for any pair of orthogonal pure states $\psi,\psi_\bot$. 
As pointed out correctly in \cite{CaiSmiBas16},
these conditions are equivalent with the positivity \eqref{Wpos} of
the transition rate operator.

\emph{Quantum State Diffusion.}
Based on decomposition \eqref{MEHfrW} of the master equation \eqref{ME}, 
a jump (piece-wise deterministic) process was constructed in ref. \cite{Dio86}, 
unraveling generic CP as well as all positive dynamics for the first time both. 
Below we concentrate on diffusive unravelings.

When the state vector $\psi$ is subject to diffusion, 
it turnes out from \eqref{MEHfrW} that a correct unraveling can be obtained if the drift velocity of $\psi$ is 
$-(i\Hfr\psi+\minihalf w)\psi$ and the matrix of diffusion is $W$. 
In a given basis numbered by lower-case Latin indices running from $1$ to $N$, the probability distribution  $p$ of the complex amplitudes $\{\psi_n,\psi_n^\star\}$
satisfies the following Fokker-Planck equation, as shown in ref. \cite{Dio88}:
\begin{eqnarray}\label{qsd1988}
\dot p&=&\frac{\partial}{\partial\psi_n}(iH_{\mathrm{fr},nm}+\minihalf w\delta_{nm})\psi_m p+c.c.\nonumber\\
&&+\frac{\partial^2 }{\partial\psi_n\partial\psi_m^\star}W_{nm}p.
\end{eqnarray}

Now we digress from the urtexts \cite{Dio86,Dio88}.
As is well known from mathematics, a Fokker-Planck equation is always equivalent
with a stochastic differential equation. Percival and Gisin considered the
CP subclass \eqref{MECP} of master equations and proposed the following
Ito-SSE \cite{QSD}:
\begin{eqnarray}\label{qsdCP}
d\psi&=&\left(-iH+\expect{\Fa}\Fa-\minihalf\Fad\Fa-\minihalf\expect{\Fad}\expect{\Fa}\right)\psi dt\nonumber\\
 &&+(\Fa-\expect{\Fa})\psi d\xi_\alpha^\star,
\end{eqnarray}
where each $\xi_\alpha$ is a standard Hermitian white-noise process with
the correlations 
\begin{equation}\label{qsdxi}
d\xi_\alpha d\xi_\beta^\star=\delta_{\alpha\beta}dt,~~~~
d\xi_\alpha d\xi_\beta=0,
\end{equation}
and with $\Ev d\xi_\alpha=0$.
One can inspect that the drift part of on the r.h.s of \eqref{qsdCP} is indeed
$-(i\Hfr\psi+\minihalf w)\psi$ while the correlation of the diffusive part yields
\begin{eqnarray}\label{WCP}
&&(\Fa-\expect{\Fa})\psi d\xi_\alpha^\star[(\Fa-\expect{\Fa})\psi d\xi_\alpha^\star]\dg\nonumber\\
=&&(\Fa-\expect{\Fa})\psi\psi\dg (\Fad-\expect{\Fad})dt=Wdt
\end{eqnarray}
as it should do.

The eqs. (\ref{qsdCP},\ref{qsdxi}) became the standard representation of QSD.
The Fokker-Planck representation \eqref{qsd1988}, valid beyond CP dynamics, went forgotten for several reasons.
First, visualization of stochastic quantum trajectories is more direct in terms
of SSEs. Second,  SSEs serve directly for Monte-Carlo simulating the solutions 
of the CP master equation \eqref{MECP}. Third, SSEs treat diffusion in finite and infinite dimensions equally well mathematically.
Nonetheless, we emphasize the merit
of the Fokker-Planck representation: it only depends on the superoperator $\Lc$
and is thus explicitly invariant against the equivalence transformations of
the  CP structure. This invariance is less obvious on the standard
eqs.  (\ref{qsdCP},\ref{qsdxi}) although one can directly prove it \cite{QSD}.

But back to the main point: the Fokker-Planck form of QSD is, as we said,
valid for all positivity-conserving master equations, even if their CP
representation does not exist. The equivalent SSE reads
\begin{equation}\label{qsdinv}
d\psi_n=-\left(iH_{\mathrm{fr},nm}+\minihalf w\delta_{nm}\right)\psi_m dt +d\chi_n
\end{equation}
where $\chi_n$ are $W$-correlated Hermitian white-noise processes:
\begin{equation}\label{qsdchinchim}
d\chi_n d\chi_m^\star=W_{nm}dt,~~~~
d\chi_n d\chi_m=0,
\end{equation}
while  $\Ev d\chi_n=0$. As we said, this form is 
representation-independent,  only depending
on the invariant operators $\Hfr$ and $W$. 
If we prefer a form with standard Hermitian white-noise processes, resembling
standard CP-QSD eqs. (\ref{qsdCP},\ref{qsdxi}), 
we decompose $W$ into the mixture of (not
necessarily normalized)  pure states orthogonal to $\psi$:
$W=\varphi_{\bot\alpha}\varphi_{\bot\alpha}\dg$. Then  the SSE reads
\begin{equation}\label{qsdnoninv}
d\psi=-(iH+\minihalf w)\psi dt +\varphi_{\bot\alpha}d\xi_\alpha^\star
\end{equation}
where the $\xi_\alpha$'s satisfy \eqref{qsdxi}. 
Advisable is that the $\varphi_{\bot\alpha}$'s be linearly independent. 
Ref. \cite{CaiSmiBas16} took the spectral decomposition of $W$ to
define  $(N-1)$ states $\{\varphi_{\bot,\alpha};\alpha=1,2,\dots,N-1\}$
orthogonal to each other (and to $\psi$).

\emph{All diffusive quantum trajectories.}
It is straightforward to find all diffusive unravelings of \emph{positive} dynamics \eqref{ME}
if we start from the invariant form of QSD (\ref{qsdinv},\ref{qsdchinchim}). Observe that the
ensemble average \eqref{unrav} of the quantum trajectories depends 
but on the Hermitian correlations of the noises, it 
is independent of $d\chi_n d\chi_m$. We can make $\chi_n$'s correlated
with themselves, generalizing \eqref{qsdchinchim}:
\begin{equation}\label{chinchim}
d\chi_n d\chi_m^\star = W_{nm}dt,~~~~
d\chi_n d\chi_m = S_{nm}dt,
\end{equation}
still we get diffusive unravelings of the same superoperator $\Lc$. While QSD
corresponds to  $S_{nm}\equiv0$, the matrix $S_{nm}$ uniquely characterizes
all diffusive unravelings, under the only constraint that the total correlation
matrix of the noises must be non-negative:
\begin{equation}\label{WS}
\left(\begin{array}{cc}d\chi d\chi\dg&d\chi d\chi\\
                                           d\chi\dg d\chi\dg&d\chi\dg d\chi
             \end{array}\right)=
\left(\begin{array}{cc}W&S\\S\dg&W\end{array}\right)\geq0.            
\end{equation}

If we start from the non-invariant representation (\ref{qsdnoninv},\ref{qsdxi}) of QSD,
the general diffusive unravelings are characterized by the correlations
\begin{equation}\label{xixi}
d\xi_\alpha d\xi_\beta^\star = \delta_{\alpha\beta}dt,~~~~
d\xi_\alpha d\xi_\beta = s_{\alpha\beta}dt,
\end{equation}
with costraint $\Vert s\Vert\leq1$. 
Note that \cite{RigMotOMa97,WisDio01} have obtained this result for the
restricted class of CP dynamics. Ref. \cite{RigMotOMa97},
citing Gisin's finding of all diffusive unravelings for 2D positive dynamics \cite{Gis90}, 
anticipated that  it might be be done in arbitrary dimensions. That's done now.
Ref. \cite{WisDio01}
had a different merit: it started from the invariant decomposition \eqref{MEHfrW}
of the dynamics \eqref{ME}. This decomposition shows explicitly that the only
constraint on the stochastic increment $d\psi$  of a diffusive quantum trajectory reads
$$
d\psi d\psi\dg = Wdt
$$ 
whereas $d\psi d\psi$ is free. It is of course possible to derive correct
SSEs in many ways, in particular representations, still the invariant method
pays in shorter calculations and in better insights.
\begin{table}
\begin{tabular}{l|cc}
                                                &positive                                              &complete positive\\
\hline
jump                 &\cite{Dio86}\cite{Dio88}\cite{Dio14}           &\cite{QT}\\
QSD                                       &\cite{Dio88,CaiSmiBas16}           &\cite{QSD}\\
all diffusive    &\cite{Gis90} [present work]                     &\cite{RigMotOMa97}\cite{WisDio01}\\
all jump           &                      ?                                       &                                ?
\end{tabular}
\caption{Summary of works on different unravelings versus positive or restricted CP dynamics.}
\end{table}

\emph{Closing remarks.}
We have not this time intended to discuss relevance and physical interpretation of unravelings. (For CP diffusive SSEs, ref. \cite{WisDio01} gave an exhaustive answer in
terms of monitoring and control of $\psi$, see \cite{Wis16}
on QSD interpretations most recently.) Role of non-CP dynamics in physics is 
not yet fully understood. The theory of their unraveling might get us 
closer to an interpretation. 

\emph{Acknowledgment.}
This work was supported by the Hungarian Scientific Research Fund under Grant No. 103917, and by the EU COST Actions MP1209 and CA15220.

\end{document}